\documentclass[a4paper,12pt]{article}
\usepackage{amsmath}
\usepackage[draft]{graphicx}
\begin{document}
\author{K.Ayuel and P.F.de Ch$\hat{a}$tel \\
Department of Physics\\
University of Amsterdam\\
Valckenierstraat 65\\
1018 XE Amsterdam\\ The Netherlands}
\title{The magnetic field generated by an electron bound in angular-momentum eigenstates}
\maketitle
\section*{Abstract}
The magnetic field generated by an electron bound in a spherically symmetric potential is calculated for eigenstates
of the orbital and total angular momentum. General expressions are
presented for the current
density in such states and the magnetic field is calculated through the vector potential,
which is obtained from the current density by direct integration. The method is applied to 
the hydrogen atom, for which we reproduce and extend known results.
\section*{I.Introduction}
Recently, Gough [1] has presented calculations of the magnetic field produced by a hydrogen atom in various angular-momentum eigenstates. The
calculations were done by solution of the differential equation for the various multipole components of the vector potential. The purpose of this 
paper is to present a method for similar calculations, where the field is calculated by a direct integration from the current density.
The multipole expansion of the latter involves only the angular part of the wave functions and we shall show
results valid for angular-momentum eigenstates irrespective of the  radial part. Our method
involves some manipulations of spherical harmonics, which may appear tedious, but provide a welcome opportunity to develop skills, which are of use in
other applications of quantum mechanics. In particular, we gain some insight in the systematics of occurrence of various multipole comportment of the
magnetic field, using identities known from the derivation of selection rules in spectroscopy.\\
The magnetic field generated by electrons occupying states in partially field shells is of interest in magnetism. Magnetic neutron scattering is mainly
due  to the interaction of the neutrons's spin with the magnetic field. The magnetic form factor [2] is said to be given by the Fourier transform
of the magnetization density of an atom. However, to avoid ambiguities 
involved in the definition of the magnetization density of an atom, it is important
to bear in mind that it is the $\mathbf{B}$ field due to unpaired electrons that is being  measured. Of course, the electronic states in a magnetic
material are different from the ones bound to free atoms. Nevertheless, the study of the latter is useful to clarify the concepts and methods involved
in the calculation of magnetic form factors. In view of their relevance
to magnetic materials, we shall discuss $\mathit{d}$  and $\mathit{ f}$ states in particular.\\

\section*{II The vector potential }
The vector potential $\mathbf{A} $ satisfies the Poisson equation\\
\begin {equation}
 \nabla^{2}\mathbf{A}=-\mu_{0}\mathbf{j}
\end{equation}
with the current density   $\mathbf{j}$ as source. In reference [1], this
differential equation has been solved for the dipole and octupole components 
of the azimuthal current density. Each Cartesian component of eq. (1) is of the
form familiar from electrostatics as one relating the potential to charge
density. Accordingly, the solution can be written in the integral form
 \\
\begin{equation}
A_{i}(\mathbf{r})=\frac{\mu_{0}}{4\pi}\int_{V^{'}}\frac{j_{i}({\mathbf{r}}^{'})dV^{'}}{R},
\end{equation}
where i stands for x,y and z, and  $R = |\mathbf{r} - {\mathbf{r}}^{'}|$ .
The factor  $\frac{1}{R}$ can be expanded [3] as 
\begin{equation}
\frac{1}{R}  = 
\frac{1}{r}\sum_{l=0}^{\infty}({\frac{r^{'}}{r}})^{l}P_{l}(\cos\alpha)
\end{equation}
for \begin{math}  r > r^{'}\end{math}, where $P_{l}$ is a Legendre  polynomial of degree $l$.This equation is also valid for $ r^{'}>r $
after interchanging $r$ with $r^{'}$. The nature of the angular-momentum 
eigenstates to be studied in the coming sections makes it convenient to work 
with spherical components. In particular, it will be seen that the current  density 
has no $r$ or $\theta$ components and therefore, since $div\mathbf{j}$ must vanish, it does not depend on $\phi$,
 i.e.,  $\mathbf{j}=j(r,\theta)\hat{\phi}$. Consequently, the
Cartesian components take the form
\begin{equation}
j_{x} = -j(r,\theta)\sin\phi ;
\end{equation}
\begin{equation}
j_{y} = j(r,\theta)\cos\phi .
\end{equation}

Substituting  eqs.(3) to (5) into eq.(2) then
gives the Cartesian components of the  vector potential,  

\begin{equation}
\begin{split}
A_{x} &= -\frac{\mu_{0}}{4\pi}  {[ }\iiint \limits_{r^{'}=0}\limits^{r}
( j(r^{'},\theta^{'})\sin\phi^{'} \frac{1}{r}\sum_{l=0}^{\infty} 
{(}\frac{r^{'}}{r})^{l} P_{l}(\cos\alpha) ) dV^{'} \\
 &+  \iiint\limits_{r^{'}=r}\limits^{\infty} ( j(r^{'},\theta^{'})\sin\phi^{'}
\frac{1}{r^{'}}\sum_{l=0}^{\infty} {(}\frac{r}{r^{'}}) ^{l} P_{l}(\cos\alpha) ) 
dV^{'} {]}
\end{split}
\end{equation}

\begin{equation}
\begin{split}
A_{y} &=\frac{\mu_{0}}{4\pi}  {[ }\iiint\limits_{r^{'}=0}\limits^{r}
( j(r^{'},\theta^{'})\cos\phi^{'} \frac{1}{r}\sum_{l=0}^{\infty} 
{(}\frac{r^{'}}{r})^{l} P_{l}(\cos\alpha) ) dV^{'} \\
 &+  \iiint\limits_{r^{'}=r}\limits^{\infty} ( j(r^{'},\theta^{'})\cos\phi^{'}
\frac{1}{r^{'}}\sum_{l=0}^{\infty} {(}\frac{r}{r^{'}}) ^{l} P_{l}(\cos\alpha) ) 
dV^{'} {]}
\end{split}
\end{equation}\\
The function $P_{l}(\cos\alpha)$ can be expanded [3] as\\
\begin{equation}
\begin{split}
P_{l}(\cos\alpha) &=P_{l}(\cos\theta)P_{l}(\cos\theta^{'})+
2\sum_{m=1}^{l}\frac{(l-m)!}{(l+m)!}P_{l}^{m}(\cos\theta)P_{l}^{m}(\cos\theta^{'}
)\\&\times\cos m(\phi-\phi^{'})\,\, ,
\end{split}
\end{equation}
where the $P_{l}^{m}$ are associated Legendre functions.
We shall decompose the current density into multipole components as 
\begin{equation}
j=\sum_{l=1,3,...}j_{l}(r)P_{l}^{1}(\cos\theta),
\end{equation}
where $j_{1}(r),j_{3}(r),...$ are the dipole, octupole,
32-pole,...component of the current density, which enables the use of
the orthogonality of the associated Legendre functions,\\
\begin{equation}
\int_{0}^{\pi}P^{m}_{p}(\cos\theta^{'})P^{m}_{q}
(\cos\theta^{'})\sin\theta^{'}d\theta^{'}=
\frac{2}{2q+1}{.}\frac{(q+m)!}{(q-m)!}\delta_{p,q}
\end{equation}

 together with the integrals
 \begin{equation}
 \int_{0}^{2\pi}\cos
 m(\phi-\phi^{'})\sin\phi^{'}d\phi^{'}=\pi\sin\phi\delta_{m,1}
 \end{equation}
 and
 \begin{equation}
 \int_{0}^{2\pi}\cos
 m(\phi-\phi^{'})\cos\phi^{'}d\phi^{'}=\pi\cos\phi\delta_{m,1} \,\, {.}
 \end{equation}
 The latter ensure that the expansion (9) is limited to $m = 1$.
  It is easily shown that for the dipole component the result is
   \begin{equation}
A_{x_{1}}= -\frac{\mu_{0}}{3} \sin\phi P^{1}_{1}(\cos\theta) {[ }
\frac{1}{r^{2}}\int_{r^{'}=0}^{r}j_{1}(r^{'}){r^{'}}^{3}dr^{'} +
 r \int_{r^{'}=r}^{\infty}j_{1}(r^{'})dr^{'} {]}
\end{equation}

\begin{equation}
A_{y_{1}}= \frac{\mu_{0}}{3} \cos\phi P^{1}_{1}(\cos\theta){[ }
\frac{1}{r^{2}}\int_{r^{'}=0}^{r}j_{1}(r^{'}){r^{'}}^{3}dr^{'} +
 r \int_{r^{'}=r}^{\infty}j_{1}(r^{'})dr^{'} {]}
\end{equation}
The dipole component of the vector potential is then of the form 
$\mathbf{A_{1}}=A_{1}(r)P^{1}_{1}(\cos\theta)\hat{\phi}$,
where

\begin{equation}
A_{1} = \frac{\mu_{0}}{3} {[}
\frac{1}{r^{2}}\int_{r^{'}=0}^{r}j_{1
}(r^{'}){r^{'}}^{3}dr^{'} +
 r \int_{r^{'}=r}^{\infty}j_{1}(r^{'})dr^{'} {]}.
\end{equation}\\
Following the same procedure, we find the further coefficients in the multipole expansion of the 
vector potential,
\begin{equation}
 \mathbf{A}=\sum_{l=1,3...}A_{l}(r)P_{l}^{1}(\cos\theta)\hat{\phi},
 \end{equation}
 to be given by

\begin{equation}
A_{l}= \frac{\mu_{0}}{2l+1} {[}
\frac{1}{r^{l+1}}\int_{r^{'}=0}^{r}j_{l}(r^{'}){r^{'}}^{l+2}dr^{'} +
 r^{l} \int_{r^{'}=r}^{\infty}\frac{j_{l}(r^{'})}{{r^{'}}^{l-1}}dr^{'}
 {]}.
\end{equation}

\section*{III  Current density in eigenstates of the angular momentum $L$ and the spin operator $S_{z}$ }

\subsection*{III.I Multipole expansion of the orbital current density in
 angular momentum eigenstates}

The orbital current density generated by an electron in the $ |n,l,m>$
angular-momentum eigenstate has been given by Gough [1] as\\
\begin{equation}
j_{\phi}^{o}=-2\mu_{B}|\psi_{nlm}|^{2}\frac{m}{r\sin\theta}\:,
\end{equation}\\
where $\mu_{B}$ is the Bohr magneton  and $ \phi$ refers to the
azimuthal component, the other components being zero, $j^{o}_{r}=j^{o}_{\theta}=0$, and \\
\begin{equation}
\psi_{nlm}= R_{nl}(r)Y_{l}^{m}(\theta,\phi)
\end{equation}\\
is the  normalized wave function . The function $j^{o}_{\phi}$ can be factorized into radial and angular parts,\\
\begin{equation}
\begin{split}
 j_{\phi}^{o} &=-2\mu_{B}\frac{R^{2}_{nl}}{r}Y_{l}^{m*} \frac{m
 Y_{l}^{m}}{\sin\theta}  \\
 &= -2\mu_{B}\frac{ R_{nl}^{2}(r)}{r}f_{lm}(\theta)
\end{split}
\end{equation}\\
where we have defined $f_{lm}(\theta) =
(-1)^mY_{l}^{-m}\frac{mY_{l}^{m}}{\sin\theta}$ . It is clear within the
formalism used in the previous section, that the various multipole components of
the current density give rise to the corresponding components of the
vector potential. Therefore, it will be convenient to carry out the
angular integrals for each multipole separately. To this end, in the present
section we shall carry out the multipole expansion of the angular part of
functions of the form (20). In doing so, we encounter the compartments
\begin{equation}
\sin\theta = P^{1}_{1}(\cos\theta)\:\:\:\:\:(dipole)\:\:;
\end{equation}
\begin{equation}
4\cos^{2}\theta\sin\theta - \sin^{3}\theta =
\frac{2}{3}P^{1}_{3}(\cos\theta)\:\:\:\:\:\:(octupole)\:\:;
\end{equation}
\begin{equation}
 8\cos^{4}\theta\sin\theta - 12\cos^{2}\theta\sin^{3}\theta +
 \sin^{5}\theta =
 \frac{8}{15}P^{1}_{5}(\cos\theta)\:\:\:(32-pole)\:\:;
\end{equation}
which we identify with the appropriate Legendre functions of $
\cos\theta$ .  To find the desired expansion coefficients for the angular
part, first the identity [4]\\
\begin{equation}
\begin{split}
\frac{m Y_{l}^{m}}{\sin\theta} &=-\frac{1}{2}\sqrt{\frac{2l+1}{2l-1}}
\big[\sqrt{(l-m-1)(l-m)}e^{-i\phi}Y_{l-1}^{ m+1} 
\\ &+\sqrt{(l+m-1)(l+m)}e^{i\phi}Y_{l-1}^{
m-1} \big]
\end{split}
\end{equation}\\
will be used and subsequently the expansion\\
\begin{equation}
Y_{l_{1}}^{m_{1}}Y_{l_{2}}^{m_{2}}=\sum_{LM}\sqrt{\frac{(2l_{1}+1)(2l_{2}+1)}
{4\pi(2L+1)}} C_{l_{1} 0 l_{2} 0}^{L 0} C_{l_{1}m_{1} l_{2}m_{2}}^{L M}
Y_{L}^{ M}\:\: ,
\end{equation}
where the Clebsch-Gordan coefficients obey the following rules :
\begin{equation}
C^{L0}_{l_{1} 0 l_{2} 0 } = 0 \:\:\:,\:\:\:unless
\:\:\:l_{1}+l_{2}+L\:is\:even\:;
\end{equation}
\begin{equation}
C^{LM}_{l_{1}m_{1} l_{2}m_{2}} =
0\:\:\:,\:\:\:\:unless\:\:\:|l_{1}-l_{2}|\leq L\leq l_{1} + l_{2} \:\:\: and
\:\: m_{1}+m_{2}=M.
\end{equation}
Collecting terms, we find 
\begin{equation}
\begin{split}
f_{lm}(\theta)= & (-1)^{m}\frac{(2l+1)}{8\pi}\sum_{L=1,3...}^{L=2l-1}
C_{l-1\, 0\:l\, 0}^{L 0} \big\{\sqrt{\frac{(l-m-1)(l-m)}{L(L+1)}}
 C_{l-1\, m+1\: l \,-m}^{L 1} \\ & -
\sqrt{\frac{(l+m-1)(l+m)}{L(L+1)}} C_{l-1\, m-1\:
l\,-m}^{L\,-1}\big\}P^{1}_{L}(\cos\theta)\:,
\end{split}
\end{equation}
where we have used the defining equations
\begin{equation}
Y_{l}^{m} = (-1)^{m}\big {[} (\frac{(2l+1)(l-m)!}{4\pi(l+m)!}\big
{]}^{\frac{1}{2}}P^{m}_{l}(\cos\theta)e^{im\phi}
\end{equation}
and 
\begin{equation}
P^{-m}_{l}(\cos\theta)=(-1)^{m}\frac{(l-m)!}{(l+m)!}P^{m}_{l}(\cos\theta)\:.
\end{equation}

Equation (28), together with the rules (26) and (27) reveals some regularities
in the occurrence of various multipole components . First, it is clear
that only the $M = 1$ components appear and secondly, $L = 1,3,..2l-1$ .
This is in accordance with the findings of Gough, in particular that in the
 $|2,1,1>$ state only the function (21) appears, whereas in $|3,2,1>$ and
 $|3,2,2> $ the functions (21) and (22). It is important to note that this
 regularity was seen to follow from the angular dependence of $j^{o}_{\phi}$
 only . This implies that the result will hold for any $p, d, f$, ect.
 states, irrespective of the radial function $R(r)$.\\

Substitution of the  Clebsch-Gordan expressions leads to Table 1.\\

Table 1. The coefficients of the $\alpha^{lm}_{L}$ in the multipole expansion of the orbital current density , see
equation (2.52)

 \begin{tabular}{l l c c c}
 \hline
 $\;$ &$\;$ &$\;$ & $\;$ &$\;$\\
 $\;$ &$\;$ &$\;$ &$L$&$\;$\\
 \cline{3-5} \\
 $l$ & $m$ & 1 & 3  &  5  \\
 $\;$ &$\;$ &$\;$ & $\;$ &$\;$\\ \hline
 $\;$ &$\;$ &$\;$ & $\;$ &$\;$\\
  3& 1& $\frac{3}{8}$ & $\frac{7}{24}$ & $\frac{5}{24}$   \\
 $\;$ &$\;$ &$\;$ & $\;$ &$\;$ \\
$\;$ &2 & $\frac{6}{8}$ & $\frac{7}{24}$ & $\frac{-4}{24}$  \\
$\;$ &$\;$ &$\;$ & $\;$ &$\;$\\
$\;$ &3 & $\frac{9}{8}$ & $\frac{-7}{24}$ &$\frac{1}{24}$   \\
$\;$ &$\;$ &$\;$ & $\;$ &$\;$ \\
\hline
$\;$ &$\;$ &$\;$ & $\;$ &$\;$\\
 2&1& $\frac{3}{8}$ & $\frac{2}{8}$ & $\;$ \\
 $\;$ &$\;$ &$\;$ & $\;$ &$\;$ \\
$\;$ & 2& $\frac{6}{8}$ & $\frac{-1}{8}$ & $\;$ \\
$\;$ &$\;$ &$\;$ & $\;$ &$\;$ \\
\hline
$\;$ &$\;$ &$\;$ & $\;$ &$\;$\\
1&1& $\frac{3}{8}$ &$\;$ & $\;$\\
$\;$ &$\;$ &$\;$ & $\;$ &$\;$ \\
\hline
\end{tabular} \\\\

 Again we find that some of the results of reference [1] are quite
general: $\mathit{p }$ states in general give rise to pure dipole fields,
$\mathit{d}$  states generate an octupole
field as well, which is opposite in sign and stronger by a factor of two for
$m_{l}=1$ than for $m_{l}=2$, etc. These and similar regularities found in Table 2 follow from the
fact that the L-th order multipole component of the current density operator is an
irreducible tensor operator of rank L.

\subsection*{III.II Multipole expansion of the spin current density for
the eigenfunctions of the spin operator $S_{z}$ }
 The current density  associated with the electron spin for eigenfunctions of
 the spin operator $S_{z}$ has been given in reference [1] as
 \begin{displaymath}
 j^{s}_{\theta}=j^{s}_{r}=0;
 \end{displaymath}
\begin{equation}
j_{\phi}^{s} = 2m_{s}\mu_{B}(\sin\theta\frac{d}{dr} +
\cos\theta\frac{d}{d\theta})\Psi\Psi^{*}\:,
\end{equation}
where
\begin{equation}
\Psi\Psi^{*} = R^{2}_{nl}Y_{l}^{m}Y_{l}^{m*} \:.
\end{equation}
 As in the previous section (III.I), eq. (25) will be used to expand $Y_{l}^{m}Y_{l}^{m*} $  as ,
\begin{equation}
Y_{l}^{m}Y_{l}^{m*} =
(-1)^{m}\sum_{L}\frac{(2l+1)}{\sqrt{4\pi(2L+1)}}C^{L 0}_{l 0 l 0}C^{L
0}_{l m l -m} Y_{L}^{ 0}(\theta,\phi) \: \: \: . 
\end{equation}
and eq. (29) will give 
$Y_{L}^{0} = (\frac{2L +1}{4\pi})^{\frac{1}{2}}P_{L }(\cos\theta) $
where $ P_{L}=P^{0}_{L}$; consequently,
\begin{equation}
Y_{l}^{m}Y_{l}^{m*} =
(-1)^{m}\frac{(2l+1)}{4\pi}\sum_{L}C^{L 0}_{l 0 l 0}C^{L
0}_{l m l -m} P_{L }(\cos\theta) \: \: \: .
\end{equation} \\

To get the multipole expansion of the spin current density given in eq. (31), we need
to express $\sin\theta P_{L}$ and $\cos\theta\frac{d}{d\theta}P_{L}$ as linear combinations 
of Legendre functions. For the former, the identity  [3],
\begin{equation}
\sin\theta P_{l}^{m} = \frac{1}{2l+1}(P_{l+1}^{m+1} - P_{l-1}^{m+1})
\end{equation}
  will be used, as a result,
\begin{equation}
\sin\theta P_{L} = \frac{1}{2L+1}(P_{L+1}^{1} - P_{L-1}^{1})\:.
\end{equation}
 For the latter, the definition of the $M=1$ associate Legendre polynomial,
\begin{equation}
\frac{d}{d\theta}P_{L}(\cos\theta)
= -P^{1}_{L}, 
\end{equation}
will be used and subsequently the identity [5]
\begin{equation}
\cos\theta P_{l}^{m} = \frac{1}{2L+1}\Big\{(l-m+1)P^{m}_{l+1} +
(l+m)P^{m}_{l-1}\Big\} ,
\end{equation}
which gives
\begin{equation}
\cos\theta P^{1}_{L} = \frac{1}{2L+1}\Big\{LP^{1}_{L+1} +
(L+1)P^{1}_{L-1}\Big\},
\end{equation}

Combining eqs. (37) and (39) we find 
\begin{equation}
\cos\theta\frac{dP_{L}}{d\theta} = -\frac{1}{2L+1}\Big\{LP^{1}_{L+1} +
(L+1)P^{1}_{L-1}\Big\} .
\end{equation}
 Substitution of eq. (34) into eq. (32) and then (32) into eq. (31) with
 the help of eqs. (36) and (40) gives
 \begin{equation}
 \begin{split}
 j^{s}_{\phi}(r,\theta) &=
 (-1)^m\mu_{B}\frac{(2l+1)}{4\pi}\sum_{L=0,2,...}^{L=2l}\frac{1}{2L+1}
 C^{L\:0}_{l\:0\:l\:0}C^{L\:0}_{l\:m\:l\:-m} \\
  & \times\Bigg\{\big( \frac{\partial R^{2}_{nl}}{\partial r} 
-\frac{L R^{2}_{nl}}{r}\big)P^{1}_{L+1} - \big(\frac{\partial 
R^{2}_{nl}}{\partial r}+(L+1)\frac{R^{2}_{nl}}{r}\big)P^{1}_{L-1}) \Bigg\}\:\:\: .
 \end{split}
 \end{equation}
  By shifting the value of  the  summation index $L$, and noticing that 
$P^{1}_{-1} = 0 $, the above
  equation can be rewritten as,
  \begin{equation}
 \begin{split}
 j^{s}_{\phi}(r,\theta) &=
 (-1)^m 2m_{s} \mu_{B}\frac{(2l+1)}{4\pi}\Bigg\{\sum_{L=1,3,..}^{L=2l+1}\Big\{\frac{1}{2L-1}
 C^{L-1\:0}_{l\:0\:l\:0}C^{L-1\:0}_{l\:m\:l\:-m}\big(\frac{\partial R^{2}_{nl}}{\partial r} \\&- 
 (L-1)\frac{R^{2}_{nl}}{r}\big)P^{1}_{L} 
 - \sum_{L=1,3,..}^{L=2l-1}\frac{1}{2L+3} C^{L+1\:0}_{l\:0\:l\:0}C^{L+1\:0}_{l\:m\:l\:-m}
 \big(\frac{\partial R^{2}_{nl}}{\partial r} 
\\&+(L+2)\frac{R^{2}_{nl}}{r}\big)P^{1}_{L} \Bigg\}.
 \end{split}
 \end{equation}

Here again we can recognize a feature of Gough's results as being the
consequence of a general rule: the highest multipole component of the
magnetic field associated with spin currents is of order $2l+1$. That
is, $s$ states generate only dipole fields, $p$ states dipole plus
octupole, ect. As we find different linear combination of
$\frac{\partial}{\partial r}R^{2}_{nl}$ and $\frac{R^{2}_{nl}}{r}$ in
the different multipole components, no general expressions can be
given for the ratio of different components. This feature is inherent
in the expression (31), which shows that the spin-current density,
unlike the orbital one, eq.(20), cannot be written as a single product
of an r-dependent and an angle-dependent factor.
\section*{IV  Current density in eigenstates of the total angular
momentum }

The eigenfunctions of the total angular momentum being superpositions
of the two eigenfunctions of $S_{z}$, the calculation of the expectation
values of the current densities involves two-component spinors. As in
the previous case, we find that the spin current density is not
factorized in radial and angular parts. However, it turns out that the
total current density  can be written as a single product of $r$- and
$\theta$-dependent functions. Therefore, we shall derive this
expression first, in section IV.I, and turn to the problem of the
multipole expansion of the radial part in section IV.II
\subsection*{IV.I The factorization of the total current density}
The eigenfunctions of the total angular momentum are of the form 
\begin{equation} \Psi = \left( \begin{array}{c}\psi_{\uparrow}(\mathbf{r}) \\ \psi_{\downarrow}(\mathbf{r}) \end{array}\right), \end{equation} 
where
\begin{equation}
\begin{array}{l}
\psi_{\uparrow}(j=l-\frac{1}{2},m_{j}=m+\frac{1}{2})=R_{nl}\sqrt{\frac{l-m}{2l+1}}Y^{m}_{l};\\
\psi_{\downarrow}(j=l-\frac{1}{2},m_{j}=m+\frac{1}{2})=-R_{nl}\sqrt{\frac{l+m+1}{2l+1}}Y^{m+1}_{l}\end{array}
\end{equation}
for the low-lying spin-orbit coupled state and 
\begin{equation}
\begin{array}{l}
\psi_{\uparrow}(j=l+\frac{1}{2},m_{j}=m+\frac{1}{2})=R_{nl}\sqrt{\frac{l+m+1}{2l+1}}Y^{m}_{l};\\
\psi_{\downarrow}(j=l-\frac{1}{2},m_{j}=m+\frac{1}{2})=R_{nl}\sqrt{\frac{l-m}{2l+1}}Y^{m+1}_{l}\end{array}
\end{equation}
for higher-lying state. The current density operators applicable to
wavefunctions of this form are represented as $2\times 2$ matrices:
\begin{equation}
 \mathbf{j}^{o}=\frac{e\hbar i}{2m_e}\left(\Psi^{*}\mathbf{\nabla}\Psi-
\Psi\mathbf{\nabla}\Psi^{*}\right) 
\end{equation}
where  $\mathbf{\nabla}$ stands for
 $\left( \begin{array}{lr} \mathbf{\nabla} & 0 \\ 0 & \mathbf{\nabla} 
\end{array}\right)$, and
\begin{equation}\mathbf{j}^{s}= -\mu_{B}\mathbf{\nabla}\times<\mathbf{\sigma}> , \end{equation}
where $\mathbf{\sigma}$ stands for the Pauli matrices,\\
\begin{equation} \sigma_{x} = \left( \begin{array}{lr} 0 & 1 \\ 1 & 0 \end{array}\right);
 \sigma_{y} = \left( \begin{array}{lr} 0 & -i \\ i & 0 \end{array}\right) ;
 \sigma_{z} = \left( \begin{array}{lr} 1 & 0 \\ 0 & -1 \end{array}\right), \end{equation}  
which can be transformed to spherical components using the
transformation matrix 
\begin{equation}
\left(\begin{array}{c}
 \sigma_{r} \\ \sigma_{\theta} \\ \sigma_{\phi}\end{array}\right)=\left(\begin{array}{ccc}
 \sin\theta\cos\phi & \sin\theta\sin\phi & \cos\theta \\
 \cos\theta\cos\phi & \cos\theta\sin\phi & -\sin\theta \\
 -\sin\phi & \cos\phi & 0     \end{array}\right)
\left( \begin{array}{c}
\sigma_{x} \\ \sigma_{y} \\ \sigma_{z} 
\end{array} \right)  
\end{equation}
giving
\begin{equation}
\begin{split} \sigma_{r} &= \left( \begin{array}{lr} \cos\theta &
\sin\theta e^{-i\phi} \\ \sin\theta e^{i\phi} & -\cos\theta \end{array}\right);
 \sigma_{\theta} = \left( \begin{array}{lr} -\sin\theta & \cos\theta
e^{-i\phi} \\ \cos\theta e^{i\phi} & \sin\theta \end{array}\right); \\& 
 \sigma_{\phi} = \left( \begin{array}{lr} 0 & -ie^{-i\phi} \\
ie^{i\phi} & 0 \end{array}\right). \end{split} \end{equation} 
As a consequence of the diagonal form of the operator (46), the orbital
current density is found to be the sum of the currents generated by the
separate spin components, given by eq. (20):
\begin{equation}
\mathbf{j}^{o}= \frac{-2\mu_{B}}{r\sin\theta}( 
m\psi_{\uparrow}^{*}\psi_{\uparrow} + (m + 
1)\psi_{\downarrow}^{*}\psi_{\downarrow} )\hat{\phi}.
\end{equation}
Substitution of the spinor components (44) gives
\begin{equation}
\begin{split}
\mathbf{j}^{o}_{j=l-\frac{1}{2}} &=\frac{-2\mu_{B}R(r)^{2}_{nl}}{r}\frac{1}{(2l+1)}\big\{
(l-m) |Y^{m}_{l}|^{2}\frac{m}{\sin\theta} \\&+ (l+m+1)|Y^{m+1}_{l}|^{2}\frac{m+1}{\sin\theta}\big\}\hat{\phi}.
\end{split}
\end{equation}
The evaluation of the spin current density is a bit more elaborate as
it requires the three components of the spin density . Taking the
expectation  values of the matrices (50) using again the spinor components
(44), we find
\begin{equation}
\begin{split}
<\sigma_{r}> &=R^{2}_{nl}\frac{1}{2l+1}\Big\{\cos\theta\big[(l-m)|Y^{m}_{l}|^{2}-
(l+m+1)|Y^{m+1}_{l}|^{2}\big] \\ &-2\sin\theta\sqrt{(l-m)(l+m+1)} Y^{m*}_{l}Y^{m+1}_{l}e^{-i\phi}\Big\};
\end{split}
\end{equation}
\begin{equation}
\begin{split}
<\sigma_{\theta}> &=R^{2}_{nl}\frac{1}{2l+1}\Big\{-\sin\theta\big[(l-m)|Y^{m}_{l}|^{2}-
(l+m+1)|Y^{m+1}_{l}|^{2}\big] \\ &-2\cos\theta\sqrt{(l-m)(l+m+1)}Y^{m*}_{l}Y^{m+1}_{l}e^{-i\phi}\Big\}; 
\end{split}
\end{equation}
\begin{equation}
<\sigma_{\phi}>=0,
\end{equation}
where we have made use of the fact that $Y^{m*}_{l}Y^{m+1}_{l}e^{-i\phi}=Y^{m}_{l}Y^{m+1\:*}_{l}e^{i\phi}$ is real.
As  $<\sigma_{\phi}>$ vanishes and  $ <\sigma_{r}>$ and $<\sigma_{\theta}>$ are independent of $\phi$ ,
eq.(47) will provide only a $\phi$ component, which depends only on $r$ and $\theta$ :
\begin{equation}
\mathbf{j}^{s}=-\frac{\mu_{B}}{r}[\frac{\partial}{\partial r}(r<\sigma_{\theta}>)-\frac{\partial}{\partial \theta}<\sigma_{r}>]\hat{\phi}.
\end{equation}
To evaluate this quantity, we require the derivative
\begin{equation}
\begin{split}
\frac{\partial}{\partial\theta}<\sigma_{r}> &=<\sigma_{\theta}>+R^{2}_{nl}\frac{1}{2l+1}\Big\{\cos\theta\big[(l-m)\frac{\partial}{\partial\theta}|Y^{m}_{l}|^{2}\\&-
(l+m+1)\frac{\partial}{\partial\theta}|Y^{m+1}_{l}|^{2}\big]
-2\sin\theta\sqrt{(l-m)(l+m+1)} \\ &\times\frac{\partial}{\partial\theta} Y^{m*}_{l}Y^{m+1}_{l}e^{-i\phi}\Big\}.
\end{split}
\end{equation}
Substituting this expression and eq.(54) into eq.(56), we find 
\begin{equation}
\begin{split}
\mathbf{j}^{s}_{j=l-\frac{1}{2}} &= -\frac{\mu_{B}}{r(2l+1)}\Big(r\frac{\partial R^{2}_{nl}}{\partial r}\bigg\{-\sin\theta\big[(l-m)|Y^{m}_{l}|^{2}\\&-
(l+m+1)|Y^{m+1}_{l}|^{2}]  -2\cos\theta\sqrt{(l-m)(l+m+1)} Y^{m*}_{l}Y^{m+1}_{l}e^{-i\phi}\Big\} 
\\&-R^{2}_{nl}\Big\{\cos\theta\big[(l-m)\frac{\partial}{\partial\theta}|Y^{m}_{l}|^{2} 
-(l+m+1)\frac{\partial}{\partial\theta}|Y^{m+1}_{l}|^{2}\big]\\& -2\sin\theta\sqrt{(l-m)(l+m+1)}\frac{\partial}{\partial\theta} Y^{m*}_{l}Y^{m+1}_{l}e^{-i\phi}\Big\}\Big)\hat{\phi}.
\end{split}
\end{equation}
To eliminate the derivatives from the second pair of braces, we make repeated use of the identities [4]
\begin{equation}
\begin{split}
\frac{\partial}{\partial \theta}Y^{m}_{l}&=m\cot\theta Y^{m}_{l}+\sqrt{(l-m)(l+m+1)}Y^{m+1}_{l}e^{-i\phi} \\
&=-m\cot\theta Y^{m}_{l}-\sqrt{(l+m)(l-m+1)}Y^{m-1}_{l}e^{i\theta}
\end{split}
\end{equation}
and we find that the awkward terms containing $\frac{1}{\sin\theta}$ in eq.(52) can be combined with
$\frac{\cos^{2}\theta}{\sin\theta}$ terms, so that the seemingly disparate angular
functions appearing in $\mathbf{j}^{o}$ and $\mathbf{j}^{s}$ provide the desired factorized from for $\mathbf{j}=\mathbf{j}^{o}+\mathbf{j}^{s}$:
\begin{equation}
\begin{split}
\mathbf{j}_{j=l-\frac{1}{2}}&=\frac{\mu_{B}}{2l+1}\big[\frac{\partial R^{2}_{nl}}{\partial r}+2(l+1)\frac{R^{2}_{nl}}{r}\big]
\Big\{\sin\theta\big[(l-m)|Y^{m}_{l}|^{2}\\&- (l+m+1)|Y^{m+1}_{l}|^{2}\big]
+2\cos\theta\sqrt{(l-m)(l+m+1)}Y^{m*}_{l}Y^{m+1}_{l}e^{-i\phi}\Big\}\hat{\phi}.
\end{split}
\end{equation}
This indeed is a product of a radial and an angular function, the former being independent of $m$. Similarly, for the case $j=l+\frac{1}{2}$ we find
\begin{equation}
\begin{split}
\mathbf{j}_{j=l+\frac{1}{2}}&=\frac{\mu_{B}}{2l+1}\big[\frac{\partial R^{2}_{nl}}{\partial r}-2l\frac{R^{2}_{nl}}{r}\big]
\Big\{\sin\theta\big[(l+m+1)|Y^{m}_{l}|^{2}-(l-m)|Y^{m+1}_{l}|^{2}\big] \\ &
-2\cos\theta\sqrt{(l-m)(l+m+1)}Y^{m*}_{l}Y^{m+1}_{l}e^{-i\phi}\Big\}\hat{\phi}.
\end{split}
\end{equation}
\subsection*{IV.II Multipole expansion of angular part.}
To find the multipole components of the total current density, as given in eqs.(60)
and (61), we again use eqs.(25) and (29). These readily yield eq.(34) and
\begin{displaymath}
Y^{m*}_{l}Y^{m+1}_{l}e^{-i\phi}=\frac{(-1)^{m+1}}{4\pi}\sum\frac{2l+1}{\sqrt{L(L+1)}}C^{L\;0}_{l0\;l0}C^{L\;1}_{l\;-m\;l\;m+1} P^{1}_{L}.
\end{displaymath}
To come to our final result, we require eqs.(36) and (39) and the identity [4]
\begin{equation}
 \sqrt{L(L+1)}C^{L\;1}_{l\;-m\;l\;m+1}=\sqrt{(l-m)(l+m+1)}(C^{L\;0}_{l\;-(m+1)\;l\;m+1}+C^{L\;0}_{l\;-m\;l\;m}).
\end{equation}
 As before, some shifting of summation indices is necessary, in order to group terms according
to the order of the Legendre function they belong to and this results in: 
\begin{equation}
\begin{split}
\mathbf{j}_{j=l-\frac{1}{2},m_{j}} &=\mu_{B}\frac{(-1)^{m}}{4\pi}\Big[\frac{\partial R^{2}_{nl}}{\partial r}+2(l+1)\frac{R^{2}_{nl}}{r}\Big]
\Big[\sum_{L=1,3,..}^{2l+1}\frac{1}{L(2L-1)}C^{L-1\;0}_{l\;0\;l\;0}\big\{
(l-m)[L \\
&-2(l+m+1)]C^{L-1\;0}_{l\;m\;l\;-m}+(l+m+1)[L-2(l-m)]C^{L-1\;0}_{l\;m+1\;l\;-(m+1)}\big\}P^{1}_{L}\\
&-\sum_{L=1,3,..}^{2l-1}\frac{1}{(L+1)(2L+3)}C^{L+1\;0}_{l\;0\;l\;0}\big\{(l-m)[L+1+2(l+m+1)]C^{L+1\;0}_{l\;m\;l\;-m} \\
&+(l+m+1)[L+1+2(l-m)]C^{L+1\;0}_{l\;m+1\;l\;-(m+1)}\big\}P^{1}_{L}\Big]\hat{\phi}.
\end{split}
\end{equation}
With similar manipulations, we get from eq.(61)
\begin{equation}
\begin{split}
\mathbf{j}_{j=l+\frac{1}{2},m_{j}} &=\mu_{B}\frac{(-1)^{m}}{4\pi}\Big[\frac{\partial R^{2}_{nl}}{\partial r}-2l\frac{R^{2}_{nl}}{r}\Big] 
\Big[\sum_{L=1,3,..}^{2l+1}\frac{1}{L(2L-1)}C^{L-1\;0}_{l\;0\;l\;0}\big\{(l+m+1)
 [L \\
 &+2(l-m)]C^{L-1\;0}_{l\;m\;l\;-m}+(l-m)[L+2(l+m+1)]C^{L-1\;0}_{l\;m+1\;l\;-(m+1)}\big\}P^{1}_{L}\\
&-\sum_{L=1,3,..}^{2l-1}\frac{1}{(L+1)(2L+3)}C^{L+1\;0}_{l\;0\;l\;0}\big\{(l+m+1)[L+1-2(l-m)]C^{L+1\;0}_{l\;m\;l\;-m}\\
&+(l-m)[L+1-2(l+m+1)]C^{L+1\;0}_{l\;m+1\;l\;-(m+1)}\big\}P^{1}_{L}\Big]\hat{\phi}.
\end{split}
\end{equation}
Equations (63) and (64) can be further reduced to :
\begin{equation}
\begin{split}
\mathbf{j}_{j=l-\frac{1}{2},m_{j}} &=\mu_{B}\frac{(-1)^{m}}{4\pi}\Big[\frac{\partial R^{2}_{nl}}{\partial r}+2(l+1)\frac{R^{2}_{nl}}{r}\Big]
\sum_{L=1,3,..}^{2l-1}\Big[\frac{1}{L(2L-1)}C^{L-1\;0}_{l\;0\;l\;0}\big\{
(l-m)[L \\
&-2(l+m+1)]C^{L-1\;0}_{l\;m\;l\;-m}+(l+m+1)[L-2(l-m)]C^{L-1\;0}_{l\;m+1\;l\;-(m+1)}\big\}\\
&-\frac{1}{(L+1)(2L+3)}C^{L+1\;0}_{l\;0\;l\;0}\big\{(l-m)[L+1+2(l+m+1)]C^{L+1\;0}_{l\;m\;l\;-m} \\
&+(l+m+1)[L+1+2(l-m)]C^{L+1\;0}_{l\;m+1\;l\;-(m+1)}\big\}\Big]P_{L}^{1}\hat{\phi}
\end{split}
\end{equation}
\begin{equation}
\begin{split}
\mathbf{j}_{j=l+\frac{1}{2},m_{j}} &=\mu_{B}\frac{(-1)^{m}}{4\pi}\Big[\frac{\partial R^{2}_{nl}}{\partial r}-2l\frac{R^{2}_{nl}}{r}\Big] 
\Big[\sum_{L=1,3,..}^{2l-1}\Big(\frac{1}{L(2L-1)}C^{L-1\;0}_{l\;0\;l\;0}\big\{(l+m+1)
 [L \\
 &+2(l-m)]C^{L-1\;0}_{l\;m\;l\;-m}+(l-m)[L+2(l+m+1)]C^{L-1\;0}_{l\;m+1\;l\;-(m+1)}\big\}\\
&-\frac{1}{(L+1)(2L+3)}C^{L+1\;0}_{l\;0\;l\;0}\big\{(l+m+1)[L+1-2(l-m)]C^{L+1\;0}_{l\;m\;l\;-m}\\
&+(l-m)[L+1-2(l+m+1)]C^{L+1\;0}_{l\;m+1\;l\;-(m+1)}\big\}\Big)P^{1}_{L} \\
&+\frac{(2l+1)^{2}}{(4l+1)(l+m+1)}C^{2l\;0}_{l\;0\;l\;0}C^{2l\;0}_{l\;m\;l\;-m}P^{1}_{2l+1}\Big]\hat{\phi};
\end{split}
\end{equation}
where we have make use of the identity [4] $(l+m+1)C^{2l\;0}_{l\;m+1\;l\;-(m+1)}=(l-m)C^{2l\;0}_{l\;m\;l\;-m}$, which follows
from the general expression for $C^{a+b\;\alpha+\beta}_{a\;\beta\;b\;\beta}$, to eliminate and reduce the highest-order term sum in eqs.(63) and (64).
These two equations may look discouraging but eminently suitable for numerical calculation of particular cases, as will be demonstrated in section V.
Also, the coefficients of the multipole expansion are universal in the sense the ones given in table 1 were found to be.
\\ \\

Table 2 The coefficients $\alpha^{jm_{j}}_{L}$ in the multipole expansion
 of the total current density, see equation (2.54).\\
 \begin{tabular}{l l c c c c }
 \hline
 $\;$ &$\;$ &$\;$ & $\;$ &$\;$ &$\;$\\
 $\;$ &$\;$ &$\;$ &$L$&$\;$ & $\;$\\
 \cline{3-5} \\
 $j=l+\frac{1}{2}$ & $m_{j}$ & 1 & 3  &  5 & 7   \\
$\;$ &$\;$ &$\;$ & $\;$ &$\;$ &$\;$\\
  \hline
  $\;$ &$\;$ &$\;$ & $\;$ &$\;$ &$\;$\\
$\frac{7}{2}$& $\frac{7}{2}$& $\frac{4}{3}$ & $-\frac{14}{33}$ & $\frac{4}{39}$ & $-\frac{5}{429}$  \\
$\;$ &$\;$ &$\;$ & $\;$ &$\;$ &$\;$\\
$\;$ &$\frac{5}{2}$ & $\frac{20}{21}$ & $\frac{10}{33}$ & $-\frac{92}{273}$& $\frac{35}{429}$   \\
$\;$ &$\;$ &$\;$ & $\;$ &$\;$ &$\;$\\
$\;$ &$\frac{3}{2}$ & $\frac{4}{7}$ & $\frac{14}{33}$ &$\frac{68}{273}$ & $-\frac{35}{143}$ \\
$\;$ &$\;$ &$\;$ & $\;$ &$\;$ &$\;$\\
$\;$ &$\frac{1}{2}$ & $\frac{4}{21}$ & $\frac{2}{11}$ &$\frac{20}{91}$ & $\frac{175}{429}$  \\
$\;$ &$\;$ &$\;$ & $\;$ &$\;$ &$\;$\\
\hline  \\ $\frac{5}{2}$ &$\frac{5}{2}$ & $\frac{9}{7}$ & $-\frac{1}{3}$ &$\frac{1}{21}$ \\
$\;$ &$\;$ &$\;$ & $\;$ &$\;$ &$\;$\\
$\;$&$\frac{3}{2}$ &$\frac{27}{35}$ &$\frac{7}{15}$ &$-\frac{5}{21}$    \\
$\;$ &$\;$ &$\;$ & $\;$ &$\;$ &$\;$\\
$\;$ &$\frac{1}{2}$ &$\frac{9}{35}$ &$\frac{4}{15}$ &$\frac{10}{21}$    \\
$\;$ &$\;$ &$\;$ & $\;$ &$\;$ &$\;$\\
\hline
$\;$ &$\;$ &$\;$ & $\;$ &$\;$ &$\;$\\
 $\frac{3}{2}$&$\frac{3}{2}$&$\frac{6}{5} $ & $-\frac{1}{5} $\\
$\;$ &$\;$ &$\;$ & $\;$ &$\;$ &$\;$\\
$\;$&$\frac{1}{2}$&$\frac{2}{5} $ & $\frac{3}{5} $ \\
$\;$ &$\;$ &$\;$ & $\;$ &$\;$ &$\;$\\
\hline
$\;$ &$\;$ &$\;$ & $\;$ &$\;$ &$\;$\\
 $\frac{1}{2}$&$\frac{1}{2}$&$1$ \\
 $\;$ &$\;$ &$\;$ & $\;$ &$\;$ &$\;$\\
\hline
\end{tabular} \\ \\

 Table 2 lists such coefficients
according to the definition
\begin{equation}
\mathbf{j}_{j=l\pm\frac{1}{2},m_{j}} =\pm\frac{\mu_{B}}{4\pi}\Big\{\frac{\partial R^{2}_{nl}}{\partial r}+[2\mp(2j+1)]\frac{R^{2}_{nl}}{r}\Big\}
\sum_{L=1,3...}^{2j}\alpha^{jm_{j}}_{L}P^{1}_{L}
\end{equation}
The notation in this expression implies what is not immediately obvious from eqs. (65) and (66): the expansion coefficients are independent of $l$. In other words: for
instances, no matter whether a $j=5/2$ sextet involves a $d$ or $f$ electron, each of the three allowed multipole components of the current density will have the same
 relative strength for different $m_{j}$ values in both cases. Again, this is a consequence of the Wigner-Eckart theorem, which is applicable, because the total current
  density has the same property as the orbital one: its multipole component of order L is an irreducible tensor operator of rank L.

\section*{V Examples }

\subsection*{V.I The magnetic field generated by the
orbital current in  the  state $|n=3,l=2,m_{l}=1>$ }

  Since the  radial part of the wave function of the state $ |3,2,1>$ is
  \begin{equation}
  R_{32}(r) = \frac{2\sqrt{30}}{1215\sqrt{a_{0}^{7}}}e^{\frac{-r}{3a_{0}}}r^{2} ,
  \end{equation}
after substitution of $R_{32}^{2} $ and $f_{21}(\theta)$ from table 1 into eq.(20) 
we get,

\begin{equation}
j_{\phi} = \frac{-16\mu_{B}}{5(3^{9})(a_{0}^{7})}e^{\frac{-2r}{3a_{0}}}r^{3}
(\frac{3}{8\pi}P_{1}^{1} +\frac{2}{8\pi}P_{3}^{1})
\end{equation}
\\
Comparing the above equation with eq.(9) gives,\\
 $j_{1}(r) = \frac{-2}{5(3^{8})(a_{0}^{7})\pi}e^{\frac{-2r}{3a_{0}}}r^{3} $ 
  and  $j_{3}(r) = \frac{-4}{5(3^{9})(a _{0}^{7})\pi}e^{\frac{-2r}{3a_{0}}}r^{3} 
$ . Substitution these js into eq.(17) and integrating using the identity [1] ,
\begin{equation}
\int r^{n}e^{-r/a}dr = 
-ae^{-r/a}\{r^{n}+nar^{n-1}+n(n-1)a^{2}r^{n-2}+...+n!a^{n}\}
\end{equation}
we have ,
\begin{equation}
\begin{split}
 A_{1} &= \frac{\mu_{0}\mu_{B}\sin\theta}{4\pi 
a_{0}^2}\big\{ e^{\frac{-2r}{3a_{0}}} (\frac{2}{(5)(3^{6})}\frac{r^{3}}{a_{0}^{3}}
+\frac{8}{(5)(3^{5})}\frac{r^{2}}{a_{0}^{2}} 
\\&+\frac{19}{(5)(3^{4})}\frac{r}{a_{0}}+ \frac{2}{9}+\frac{2}{3}\frac{a_{0}}{r}+
\frac{a_{0}^{2}}{r^{2}})-\frac{a_{0}^{2}}{r^{2}}\big \}
\end{split}
\end{equation}  

\begin{equation}
\begin{split}
 A_{3} &= 
\frac{27\mu_{0}\mu_{B}(4\cos^{2}\theta\sin\theta-\sin^{3}\theta)}{2\pi
a_{0}^2}\big\{ e^{\frac{-2r}{3a_{0}}} (\frac{1}{(5)(3^{9})}\frac{r^{3}}{a_{0}^{3}} 
+\frac{4}{(5)(3^{8})}\frac{r^{2}}{a_{0}^{2}}\\&+\frac{4}{(5)(3^{6})}\frac{r}{a_{0}}+ \frac{2}{3^{5}}+\frac{4}{3^{4}}\frac{a_{0}}{r}+
\frac{2}{9}\frac{a_{0}^{2}}{r^{2}} + \frac{2}{3}\frac{a_{0}^{3}}{r^{3}}+ 
\frac{a_{0}^{4}}{r^{4}} )- \frac{a_{0}^{4}}{r^{4}}\big \} \\
\end{split}
\end{equation}

 The magnetic filed $\mathbf{B} $ = curl $\mathbf{A}$  has two components 
[1],
$
B_{r} = \frac{1}{r\sin\theta}\frac{\partial}{\partial\theta}(\sin\theta 
A_{\phi})
$, and 
 $
 B_{\theta} = -\frac{1}{r}\frac{\partial}{\partial r }(rA_{\phi})
$.Evaluation of these  components for the $ A_{1}$ and $A_{3} $ results in 
the magnetic fields , $B_{r_{1}}$ , $B_{r_{3}}$, $B_{\theta_{1}} $  and 
$B_{\theta_{3}} $ where,
\begin{equation}
\begin{split}
B_{r_{1}} &= \frac{\mu_{0}\mu_{B}\cos\theta}{2\pi a_{0}^3}\big\{ 
e^{\frac{-2r}{3a_{0}}} (\frac{2}{(5)(3^{6})}\frac{r^{2}}{a_{0}^{2}}
+\frac{8}{(5)(3^{5})}\frac{r}{a_{0}}
\\&+\frac{19}{(5)(3^{4})}+ 
\frac{2}{9}\frac{a_{0}}{r}+\frac{2}{3}\frac{a_{0}^{2}}{r^{2}}+
\frac{a_{0}^{3}}{r^{3}})-\frac{a_{0}^{3}}{r^{3}}\big \}
\end{split}
\end{equation}
\begin{equation}
\begin{split}
B_{r_{3}} &  = 
\frac{54\mu_{0}\mu_{B}(5\cos^{3}\theta-3\cos\theta)}{\pi
a_{0}^3}\big\{ e^{\frac{-2r}{3a_{0}}} (\frac{1}{(5)(3^{9})}\frac{r^{2}}{a_{0}^{2}}
 +\frac{4}{(5)(3^{8})}\frac{r}{a_{0}} \\
 & +\frac{4}{(5)(3^{6})}+ 
\frac{2}{3^{5}}\frac{a_{0}}{r}+\frac{4}{3^{4}}\frac{a_{0}^{2}}{r^{2}}+
\frac{2}{9}\frac{a_{0}^{3}}{r^{3}} + \frac{2}{3}\frac{a_{0}^{4}}{r^{4}}+
\frac{a_{0}^{5}}{r^{5}} )- \frac{a_{0}^{5}}{r^{5}}\big \} \\
\end{split}
\end{equation}

\begin{equation}
\begin{split}
B_{\theta_{1}} & = \frac{\mu_{0}\mu_{B}\sin\theta}{4\pi
a_{0}^3}\big\{ e^{\frac{-2r}{3a_{0}}} (\frac{4}{(5)(3^{7})}\frac{r^{3}}{a_{0}^{3}}
+\frac{8}{(5)(3^{6})}\frac{r^{2}}{a_{0}^{2}} \\
 & +\frac{14}{(5)(3^{5})}\frac{r}{a_{0}}+ \frac{22}{5(3^{4})}+\frac{2}{9}\frac{a_{0}}{r}+
\frac{2}{3}\frac{a_{0}^{2}}{r^{2}} +\frac{a_{0}^{3}}{r^{3}} 
)-\frac{a_{0}^{3}}{r^{3}}\big \}\\
\end{split}
\end{equation}

\begin{equation}
\begin{split}
 B_{\theta_{3}} & =
\frac{81\mu_{0}\mu_{B}(4\cos^{2}\theta\sin\theta-\sin^{3}\theta)}{2\pi
a_{0}^3}\big\{ e^{\frac{-2r}{3a_{0}}} (\frac{2}{(5)(3^{11})}\frac{r^{3}}{a_{0}^{3}}
 +\frac{4}{(5)(3^{10})}\frac{r^{2}}{a_{0}^{2}} \\
 & +\frac{4}{(5)(3^{8})}\frac{r}{a_{0}}+ 
\frac{4}{(5)(3^{6})}+\frac{2}{3^{5}}\frac{a_{0}}{r}+
\frac{4}{3^{4}}\frac{a_{0}^{2}}{r^{2}} + \frac{2}{9}\frac{a_{0}^{3}}{r^{3}}+
\frac{2}{3}\frac{a_{0}^{4}}{r^{4}}+\frac{a_{0}^{5}}{r^{5}} )- 
\frac{a_{0}^{5}}{r^{5}}\big \} \\
\end{split}
\end{equation} 
This example serves as a verification that our method reproduces the results of ref.1, which it does. 
 Figure 1 illustrates the result in terms of field lines.

\subsection*{IV.II  The magnetic field generated by the state $|n=3,l=2,j=3/2,
m_{j}=3/2> $ }
Substitution of the  radial wave function $R_{32}$ of the previous example and
 the coefficients $\alpha^{jm_{j}}_{L}$ of table 2 into eq.(67) gives
 
\begin{equation}
j_{\frac{3}{2}\frac{3}{2}}=\frac{4\mu_{B}r^{3}e^{\frac{-2r}{3a_{0}}}}{\pi5(3^{10})a_{0}^{8}}
(15a_{0}-r)(\frac{-6}{5}P^{1}_{1}+\frac{7}{35}P^{1}_{3})
\end{equation}
Following the same procedure of the above example,the expressions for the 
vector potential and for the magnetic are,
\begin{equation}
\begin{split}
 A_{1} &= \frac{3\mu_{0}\mu_{B}\sin\theta}{10\pi 
a_{0}^2}\big\{ e^{\frac{-2r}{3a_{0}}} (\frac{4}{(5)(3^{8})}\frac{r^{4}}{a_{0}^{4}}
+\frac{2}{(5)(3^{4})}\frac{r^{2}}{a_{0}^{2}} 
+\frac{2}{(5)(3^{2})}\frac{r}{a_{0}}+ \frac{2}{9} \\
&+\frac{2}{3}\frac{a_{0}}{r}+
\frac{a_{0}^{2}}{r^{2}})-\frac{a_{0}^{2}}{r^{2}}\big \}\\
\end{split}
\end{equation}

\begin{equation}
\begin{split}
 A_{3} & = 
\frac{-27\mu_{0}\mu_{B}(4\cos^{2}\theta\sin\theta-\sin^{3}\theta)}{20\pi
a_{0}^2}\big\{ e^{\frac{-2r}{3a_{0}}} (\frac{4}{(5)(3^{8})}\frac{r^{2}}{a_{0}^{2}}
+\frac{4}{(5)(3^{6})}\frac{r}{a_{0}}\\&+ \frac{2}{3^{5}} +\frac{4}{3^{4}}\frac{a_{0}}{r}+
\frac{2}{9}\frac{a_{0}^{2}}{r^{2}} + \frac{2}{3}\frac{a_{0}^{3}}{r^{3}}+ 
\frac{a_{0}^{4}}{r^{4}}-\frac{2}{(5)(3^{10})}\frac{r^{4}}{a_{0}^{4}} )- \frac{a_{0}^{4}}{r^{4}}\big \} 
\end{split}
\end{equation}
\begin{equation}
\begin{split}
B_{r_{1}} &= \frac{3\mu_{0}\mu_{B}\cos\theta}{10\pi a_{0}^3}\big\{ 
e^{\frac{-2r}{3a_{0}}} (\frac{2}{(5)(3^{4})}\frac{r}{a_{0}}
+\frac{2}{(5)(3^{2})}+ 
\frac{2}{9}\frac{a_{0}}{r}
+\frac{2}{3}\frac{a_{0}^{2}}{r^{2}}\\
&+\frac{a_{0}^{3}}{r^{3}}-\frac{4}{(5)(3^{8})}\frac{r^{3}}{a_{0}^{3}})-\frac{a_{0}^{3}}{r^{3}}\big \}\\
\end{split}
\end{equation}
\begin{equation}
\begin{split}
B_{r_{3}} &  = 
\frac{-27\mu_{0}\mu_{B}(5\cos^{3}\theta-3\cos\theta)}{5\pi
a_{0}^3}\big\{ e^{\frac{-2r}{3a_{0}}} (\frac{4}{(5)(3^{8})}\frac{r}{a_{0}} 
 +\frac{4}{(5)(3^{6})}\\&+ 
\frac{2}{3^{5}}\frac{a_{0}}{r}+\frac{4}{3^{4}}\frac{a_{0}^{2}}{r^{2}}+\frac{2}{9}\frac{a_{0}^{3}}{r^{3}} + \frac{2}{3}\frac{a_{0}^{4}}{r^{4}}+
\frac{a_{0}^{5}}{r^{5}}-\frac{2}{(5)(3^{10})}\frac{r^{3}}{a_{0}^{3}} )- \frac{a_{0}^{5}}{r^{5}}\big \} \\
\end{split}
\end{equation}

\begin{equation}
\begin{split}
B_{\theta_{1}} & = \frac{3\mu_{0}\mu_{B}\sin\theta}{10\pi
a_{0}^3}\big\{ e^{\frac{-2r}{3a}} (\frac{4}{(5)(3^{8})}\frac{r^{3}}{a_{0}^{3}}
+\frac{4}{(5)(3^{5})}\frac{r^{2}}{a_{0}^{2}} 
 +\frac{2}{(5)(3^{3})}\frac{r}{a_{0}}\\
 &+ \frac{8}{5(3^{3})}+\frac{2}{9}\frac{a_{0}}{r}+
\frac{2}{3}\frac{a_{0}^{2}}{r^{2}} +\frac{a_{0}^{3}}{r^{3}} 
-\frac{8}{(5)(3^{9})}\frac{r^{4}}{a_{0}^{4}})-\frac{a_{0}^{3}}{r^{3}}\big \}\\
\end{split}
\end{equation}

\begin{equation}
\begin{split}
 B_{\theta_{3}} & =
\frac{-81\mu_{0}\mu_{B}(4\cos^{2}\theta\sin\theta-\sin^{3}\theta)}{20\pi
a_{0}^3}\big\{ e^{\frac{-2r}{3a}} (\frac{2}{3^{11}}\frac{r^{3}}{a_{0}^{3}}
 +\frac{8}{(5)(3^{10})}\frac{r^{2}}{a_{0}^{2}} 
  \\&+\frac{4}{(5)(3^{8})}\frac{r}{a_{0}}\frac{4}{(5)(3^{6})}+\frac{2}{3^{5}}\frac{a_{0}}{r}+
\frac{4}{3^{4}}\frac{a_{0}^{2}}{r^{2}} + \frac{2}{9}\frac{a_{0}^{3}}{r^{3}}+
\frac{2}{3}\frac{a_{0}^{4}}{r^{4}}+\frac{a_{0}^{5}}{r^{5}}\\&-\frac{4}{(5)(3^{12})}\frac{r^{4}}{a_{0}^{4}} )- 
\frac{a_{0}^{5}}{r^{5}}\big \} \\
\end{split}
\end{equation} 
\\ 

Figure 2 is a graphical representation of the magnetic field in this case. The pattern is much simpler than the previous one; it
 is reminiscent of the field of a simple current loop. The same holds for the case $j=\frac{5}{2}$  $m_{j}=\frac{5}{2}$, as seen in fig. 3 . This becomes
 understandable if one notices, using the coefficients of table 2, that $j(r,\theta)$ is proportional to $\sin^{3}\theta$ for $j=m_{j}=\frac{3}{2}$
 and to $\sin^{5}\theta$ for $j=m_{j}=\frac{5}{2}$. Thus, with increasing $j$, the current tends to be confined in the 'equatorial' plane. Comparing figs. 3 and 4,
 one realizes the limits of the 'universality' of the relations summarized in table 2. Although in these two cases the $\theta$-dependence of $j$ is identical, the different
 r-dependences influence the resulting pattern of field lines in a qualitative way.

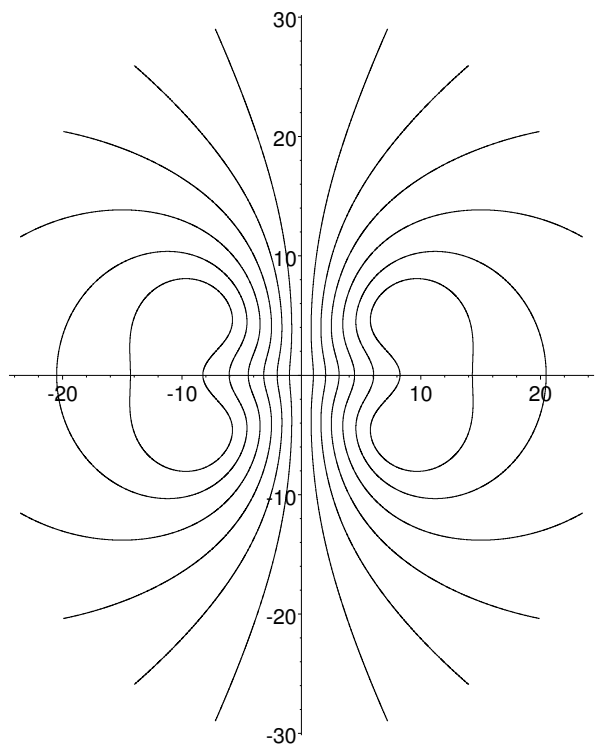
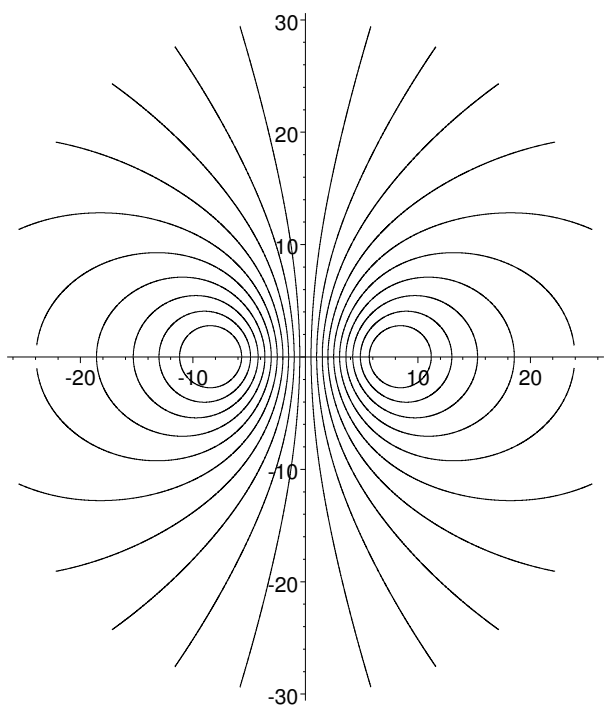
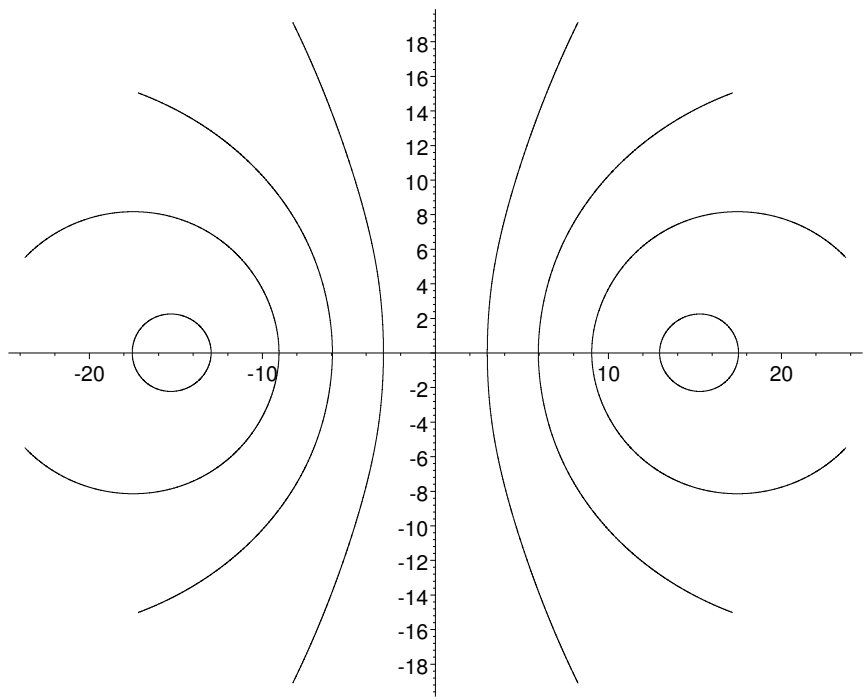
\end{document}